# Black brookite rich in oxygen vacancies as an active photocatalyst for $CO_2$ conversion: experiments and first-principles calculations

Masae Katai[1,2], Parisa Edalati[3], Jacqueline Hidalgo-Jimenez[1,4], Yu Shundo[1,2], Taner Akbay[5], Tatsumi Ishihara[1,2,6], Makoto Arita[7], Masayoshi Fuji[3,8] and Kaveh Edalati[1,2,]*

[1] WPI, International Institute for Carbon-Neutral Energy Research (WPI-I2CNER), Kyushu University, Fukuoka 819-0395, Japan
[2] Mitsui Chemicals, Inc. - Carbon Neutral Research Center (MCI-CNRC), Kyushu University, Fukuoka 819-0395, Japan
[3] Department of Life Science and Applied Chemistry, Nagoya Institute of Technology, Tajimi 507-0071, Japan
[4] Graduate School of Integrated Frontier Sciences, Department of Automotive Science, Kyushu University, Fukuoka, Japan
[5] Materials Science and Nanotechnology Engineering, Yeditepe University, Istanbul 34755, Turkey
[6] Department of Applied Chemistry, Faculty of Engineering, Kyushu University, Fukuoka 819-0395, Japan
[7] Department of Materials Science and Engineering, Faculty of Engineering, Kyushu University, Fukuoka 819-0395, Japan
[8] Advanced Ceramics Research Center, Nagoya Institute of Technology, Tajimi 507-0071, Japan

**Abstract**
Photocatalytic $CO_2$ conversion is a clean technology to deal with $CO_2$ emissions, and titanium oxide ($TiO_2$) polymorphs are the most investigated photocatalysts for such an application. In this study, black $TiO_2$ brookite is produced by a high-pressure torsion (HPT) method and employed as an active photocatalyst for $CO_2$ conversion. Black brookite with a large concentration of lattice defects (vacancies, dislocations and grain boundaries) showed enhanced light absorbance, narrowed optical bandgap and diminished recombination rate of electrons and holes. The photocatalytic activity of the black oxide for $CO_2$ conversion was higher compared to commercial brookite and benchmark P25 catalyst powders. First-principles calculations suggested that the presence of oxygen vacancies in black brookite is effective not only for reducing optical bandgap but also for providing active sites for the adsorption of $CO_2$ on the surface of $TiO_2$.

*Corresponding Author (E-mail: kaveh.edalati@kyudai.jp; Tel: +81-92-802-6744)

# 1. Introduction

Carbon dioxide ($CO_2$) is a greenhouse gas that is mainly emitted from burning fossil fuels in chemical processes and other human activities [1,2]. Global warming is a destructive result of such emissions which have negatively influenced the life cycle on the Earth [1,2]. Photocatalytic $CO_2$ conversion on a semiconducting catalyst under solar irradiation converts the anthropogenic $CO_2$ gas to CO and/or other value-added components and fuels (CO is an intermediate product in the carbene $CO_2$ conversion pathway, while CO is not formed during the formaldehyde and glyoxal pathways) [1,2]. In photocatalytic $CO_2$ conversion, the semiconductor should have desired features for absorbing the light, transferring electrons from the valence band to the conduction band, migrating the excitons to the surface and promoting the reduction reactions by electrons and oxidation reactions by holes [1,2]. Titanium oxide ($TiO_2$) has widely been used for photocatalytic $CO_2$ conversion and other solar-driven technologies [3]. This material can be present in three main types of polymorphs such as anatase, rutile and brookite, although the oxide has also some high-pressure polymorphs such as columbite and baddeleyite [4]. Although anatase and rutile have been widely investigated for photocatalytic $CO_2$ conversion [3,4], studies on the use of brookite for $CO_2$ reduction are quite limited [5-7].

Investigation on the photocatalytic activity of brookite suggested that a high degree of openness in brookite structure leads to acceleration of the electron-hole separation and migration which makes this polymorph promising for photocatalysis [4-6]. Furthermore, the theoretical calculations showed that the level of the conduction band in brookite is higher than rutile and anatase and surface activity is high, making this polymorph an appropriate candidate for photocatalytic activity [8]. However, the activity of this phase for photocatalytic $CO_2$ conversion is still low for practical applications due to the large bandgap of 3.1 eV and the high recombination rate of electrons and holes [3,5]. There have been various attempts to enhance the efficiency of brookite polymorph for photocatalytic $CO_2$ conversion. Formation of heterojunctions with different semiconductors such as $Cu_xS$ [9], $CeO_2$ [10] and $g-C_3N_4$ [11] and also with other $TiO_2$ polymorphs such as anatase [12,13] is the most reported method. Producing nanosheets [5], introducing exposed crystal face-controlled nanorods [14], surface decoration by co-catalysts [15], shape controlling by amino acids [16] and doping with Cu [17] are some other reported methods to enhance the photocatalytic $CO_2$ conversion by brookite.

Motivated by earlier studies on enhanced light absorbance and photocatalytic activity of black oxides [18,19], one may expect that producing black brookite can be effective in increasing the photocatalytic activity for $CO_2$ conversion. Black titania, containing large fractions of defects and oxygen vacancies, can be produced by reducing $TiO_2$ using hydrogen [19] or aluminum [18]. Another method to produce black oxides is mechanical treatment by severe plastic deformation methods such as high-pressure torsion (HPT) which was successfully used in producing black $ZrO_2$ [20].

In this study, black brookite with a large fraction of defects and vacancies was produced as an active photocatalyst for $CO_2$ conversion. Black brookite, which was produced by mechanical treatment via the HPT method, showed better light absorbance, narrower bandgap, less electron-hole recombination and finally higher photocatalytic $CO_2$ conversion compared to initial brookite powder. It was found via first-principles calculations that the high photocatalytic activity of black brookite is not only due to large light absorbance but also due to the role of oxygen vacancies as active sites for the adsorption of $CO_2$ to the surface of $TiO_2$.

## 2. Experimental procedure
### 2.1. Sample synthesis

Brookite (99.99%) powder, with the morphology shown in Fig. S1, was purchased from Kojundo Chemical Company, Japan. To produce black brookite, ~280 mg of the initial brookite $TiO_2$ powder was first pressed under 380 MPa pressure to produce a disc-shaped pellet with a 10 mm diameter. The pellet was processed by HPT anvils (having a 10 mm diameter hole with 0.25 mm depth at the center) under 6 GPa with a 1 rpm rotation rate for 1 turn at 300, 373 and 473 K. The schematic diagram of HPT is shown in Fig. 1a and detailed information about this method can be found in the literature [21,22]. The appearance of four samples, including the initial sample and the HPT-processed ones, is shown in Fig. 1b. The color of the samples changed from white for the initial powder to light grey, dark grey and black for samples processed at 300, 373 and 473 K, respectively. These color changes from light to dark can be considered as a piece of evidence for the formation of oxygen vacancies in the samples as reported in some other materials [23,24]. The four samples after the synthesis were characterized by various methods.

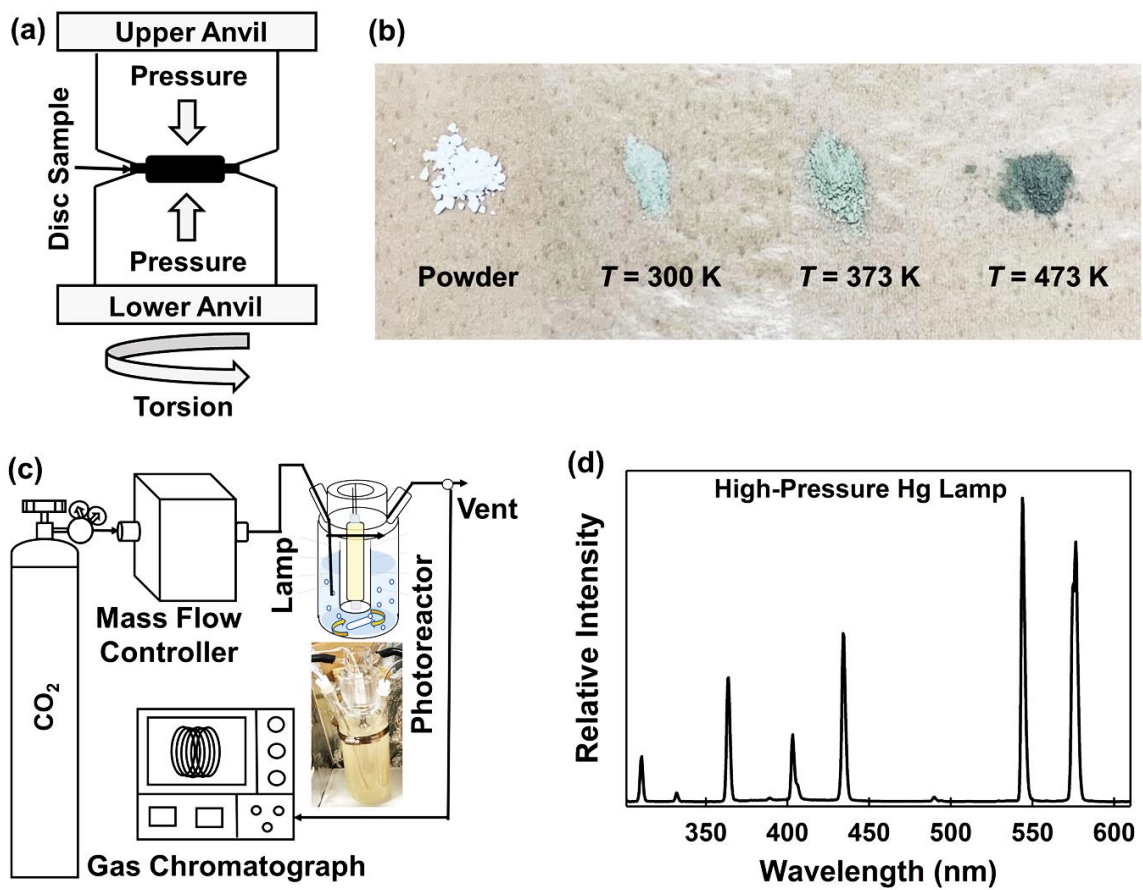

Fig. 1. (a) Schematic diagram of HPT method, (b) appearance of brookite samples before HPT and after HPT at 300, 373 and 473 K, (c) schematic diagram of photocatalytic $CO_2$ conversion test, and (d) spectral composition of mercury light source.

## 2.2. Characterizations

The crystal structure was examined by X-ray diffraction (XRD) with Cu Kα radiation and an X-ray wavelength of $\lambda = 0.1542$ nm, and by Raman spectroscopy with a laser wavelength of $\lambda = 532$ nm.

Particle size was examined by scanning electron microscopy (SEM) at an electron acceleration voltage of 15 keV.

Microstructure was investigated by transmission electron microscopy (TEM) via analyzing the bright-field (BF), dark-field (DF) and high-resolution (HR) images as well as selected area electron diffraction (SAED) and fast Fourier transform (FFT) diffractograms at 200 keV. To perform the TEM analysis, samples were crushed in ethanol and dispersed on carbon grids.

The oxidation state of elements and the level of the top of the valence band were investigated by X-ray photoelectron spectroscopy (XPS) using Mg Kα radiation with an X-ray energy of 1253.6 eV.

The recombination rate of electrons and holes was examined using steady-state photoluminescence emission spectroscopy with a laser light source of $\lambda = 325$ nm wavelength.

Oxygen vacancy formation was investigated by electron paramagnetic resonance (EPR) with a microwave source of 9.4688 GHz frequency.

The light absorbance was evaluated by UV-vis diffuse reflectance spectroscopy, and the bandgap was determined by the Kubelka-Munk theory.

## 2.3. Photocatalytic $CO_2$ conversion test

The efficiency of samples towards photocatalytic $CO_2$ conversion was investigated in a continuous quartz reactor as shown in Fig. 1c. To prepare the reaction media, 100 mg of samples were dispersed in 500 mL water with 1 M $NaHCO_3$. $CO_2$ gas was injected continuously with a 30 mL/min flow rate into the liquid of the reactor using a tube connected to a $CO_2$ gas cylinder. The reaction temperature was kept constant at 288 K using a water chiller. The photocatalytic reaction was performed under a high-pressure mercury light source (Sen Lights continuous flow Corporation, HL400BH-8, 400 W) located inside the reactor with an intensity of 0.5 W/cm$^2$. The spectral composition of the mercury light source can be observed in Fig. 1d. Before starting the reaction under light, the experiment was conducted in the dark for 2 h as a blank test. The produced gases during the reaction, were directed into a gas chromatograph (Shimadzu GC-8A, Ar Carrier) using a tube connected to the top of the reactor. To prevent the gas accumulation, another outlet flow gas also acted as a vent. The gas chromatograph was equipped with a flame ionization detector with a methanizer (Shimadzu MTN-1) and a thermal conductivity detector to determine the amount of produced CO, $CH_4$, $O_2$ and $H_2$.

## 3. Calculation procedure

First-principles calculations were conducted to achieve the band structure and the density of states (DOS) as well as to examine the adsorption of $CO_2$ molecules onto the surface of brookite in the absence and presence of oxygen vacancies. The plane-wave density functional theory (DFT) calculations were carried out using the Vienna Ab-initio Simulation Package (VASP) [25,26]. The Perdew-Burke-Ernzerhof (PBE) exchange-correlation functionals together with the projected-augmented-wave (PAW) [27] pseudopotentials were utilized throughout the calculations. The onsite Coulomb interactions were corrected by employing the GGA+U method for local correlations in the d orbitals by adding a Hubbard U parameter of 8 eV for titanium. The PAW pseudopotentials included 12 valence electrons for titanium, 6 for oxygen, 4 for carbon, and 1 for hydrogen. The wave functions were calculated with a cutoff energy of 520 eV, and the Brillouin zone was sampled using a k-point grid of 6×6×4 based on the crystal structure. The optimization process involved using the conjugate gradient algorithm, with the electronic convergence criteria of $1.0×10^{-6}$ eV and the ionic convergence criteria of $1.0×10^{-5}$ eV/Å$^2$. The graphical representation was done using the VESTA software [28].

For the investigation of the effect of oxygen vacancies on band structure, a supercell of columbite with 8 titanium atoms and 16 oxygen atoms was simulated and subsequently 1 out of 16 oxygen atoms was removed to simulate a structure with 6% oxygen vacancies. The pristine and oxygen-deficient supercells were optimized before the self-consistent field (SCF) calculations and employed for the DOS analyses.

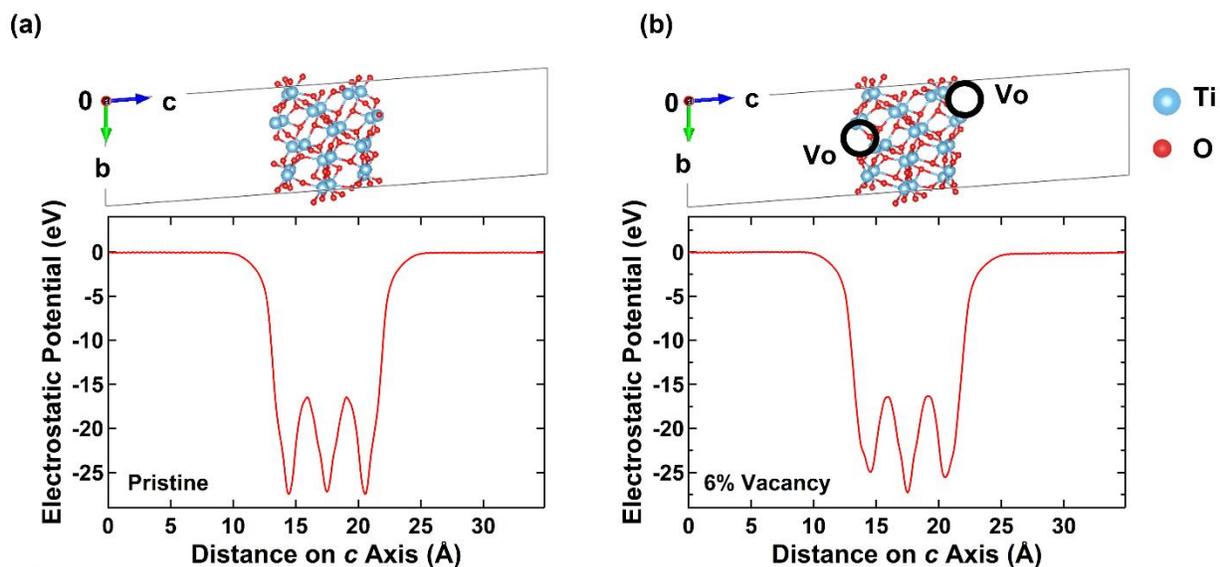

Fig. 2. Structure of brookite with (111) surface slab exposed to vacuum and corresponding potential energy calculation for (a) pristine and (b) oxygen-deficient brookite with 6% oxygen vacancy.

For examination of the effect of oxygen vacancies on the adsorption of $CO_2$ to the surface of brookite, the (111) surface slab of brookite was generated using the Surfaxe code [29] with 10 Å of columbite thickness and 30 Å of vacuum thickness on the lower and upper sides of brookite with zero net electrostatic dipole moments. The (111) slab was selected because it was reported as a stable atomic plane of brookite [30] and highly active for $TiO_2$ photocatalysts [31]. The surface termination consists of four- and five-coordinated titanium atoms and three and one-coordinated oxygen atoms. The $CO_2$ molecule was set over a four-coordinated titanium atom. To simulate the oxygen vacancies, one oxygen atom was removed from the upper side and another one from the lower side of brookite. The pristine and oxygen-deficient structures were optimized using the electronic convergence criteria of $1.0×10^{-6}$ eV and the ionic relaxation force criteria of $1.0×10^{-5}$ eV/Å. Both Brillouin zones were sampled with a 6×6×1 k-point grid. Fig. 2 shows the modeled structure and the calculated potentials for (a) pristine and (b) oxygen-deficient models, confirming their non-polar features. After the initial geometry optimization, a $CO_2$ molecule was placed above an oxygen atom and corresponding optimization was performed. The binding energy of $CO_2$ to brookite was calculated as the energy of the brookite plus $CO_2$ after the adsorption of $CO_2$ minus the energy of brookite and $CO_2$ before the adsorption of $CO_2$.

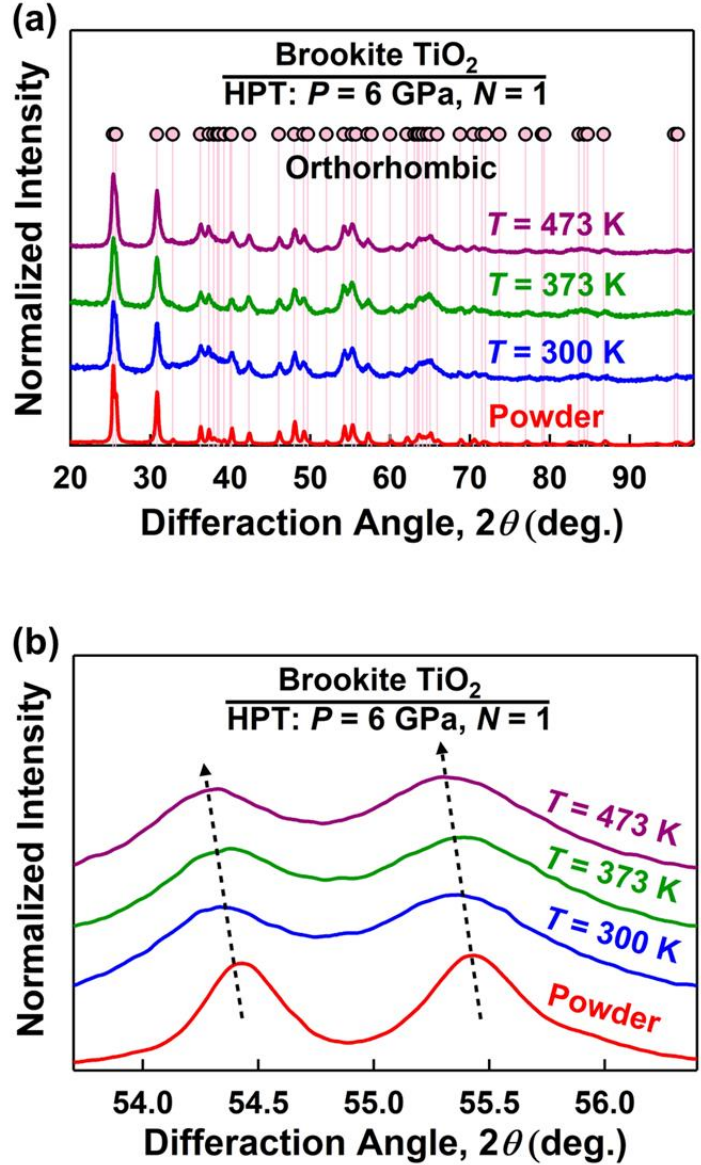

Fig. 3. (a) XRD spectra and (b) magnified XRD profile at 53.5-56.5° for brookite samples before and after HPT at 300, 373 and 473 K.

## 4. Results

Fig. 3a shows the XRD profile of the initial brookite powder and HPT samples processed at 300, 373 and 473 K. All samples have an orthorhombic structure with the Pbca space group (*a*

= 5.163 nm, $b$ = 9.159 nm, $c$ = 5.439 nm, $α$ = 90°, $β$ = 90°, $γ$ = 90°) in agreement with JCPDF 76-1934 card for brookite. A peak broadening is observed for samples processed by HPT compared to the initial powder, as shown more clearly in Fig. S2. Peak broadening is an indication of the lattice strain due to the formation of linear and planar defects after HPT processing. Fig. 3b shows the magnified XRD profiles of the four samples for the peaks located at 54.5-55.5°. This figure illustrates that a systematic peak shift to the lower diffraction angles happens by the HPT process and the shift becomes more intense with increasing the processing temperature. This peak shift corresponds to the lattice expansion due to the presence of oxygen vacancies in HPT-processed samples [23,32]. Fig. 4a shows the Raman spectra of four samples, indicating the presence of a single orthorhombic phase in agreement with XRD results. The Raman spectroscopy of brookite includes 16 peaks as shown in Fig. 4a. It includes seven peaks in $A_{1_2}$ mode located at 124, 151, 192, 244, 410, 544, 636 cm$^{-1}$, four peaks in $B_{1_g}$ mode located at 211, 318, 415, 500 cm$^{-1}$ and four peaks in $B_{2_g}$ mode located at 365, 393, 460, 581 cm$^{-1}$ [5,7,33]. Fig. 4b is the magnified Raman spectra of samples for the peaks located around 154 cm$^{-1}$. A shift can be observed for these peaks. This $A_{1_2}$ mode Raman peak shift, which increases by raising the temperature during the HPT process, was attributed to the formation of oxygen vacancies in Ref. [34] in agreement with the color of samples and XRD results.

Statistical analysis of brookite samples by SEM indicates that the average particle size increases from 2.63 μm for the powder to 7.53, 7.64 and 8.15 μm for HPT samples processed at 300, 373 and 473 K, respectively. Fig. S1 shows some representative SEM micrographs for (a) the initial brookite powder and HPT samples processed at (b) 300 K, (c) 373 K and (d) 473 K. The specific surface area of each sample can be calculated by considering the particle size values, indicating that the specific surface area decreases from 0.55 m$^2$g$^{-1}$ for the initial powder to 0.19, 0.19 and 0.18 for HPT samples processed at 300, 373 and 473 K, respectively. Reduction of the specific surface area is a general feature of catalysts synthesized under high pressure using the HPT method [20-24].

Examination of microstructures by TEM indicates that the formation of nanograins after HPT processing. Fig. S3 shows the BF, DF and SAED images of (a-c) the initial brookite powder, and HPT samples processed at (d-f) 300 K, (g-i) 373 K and (j-l) 473 K, respectively. Since, white regions in DF images are representative of grains, so existence of nanograins in HPT samples is observed clearly. The ring pattern of the SAED image for HPT samples also confirms the presence of nanograins with random misorientations. Fig. 5 shows the HR TEM images of (a) initial brookite powder and HPT samples processed at (b) 300 K, (c) 373 K and (d) 473 K. The existence of nanograin boundaries, which are evident after HPT processing, may contribute to the easy separation of electrons and holes and their migration [35]. Formation of dislocations and nanograin boundaries after HPT processing is shown more clearly in Fig. 6 for processing temperatures of (a,b) $T$ = 300 K, (c,d) $T$ = 373 K and (e,f) $T$ = 473 K. Dislocations are also considered as effective defects to improve the photocatalytic properties and activity [35].

Fig. 7a and 7b show the XPS results for the oxidation state of Ti 2p and O 1s for four samples, confirming the existence of Ti$^{4+}$ cations and O$^{2-}$ anions, respectively. By considering the C 1s XPS peak at 284.8 eV, the O 1s peak shifts to higher levels of energy after HPT processing (Fig. 7b). This peak shift, which is more significant at higher HPT processing temperatures, can be attributed to the formation of oxygen vacancies after HPT processing [36] in agreement with XRD and Raman results and also in good consistency with the color of samples.

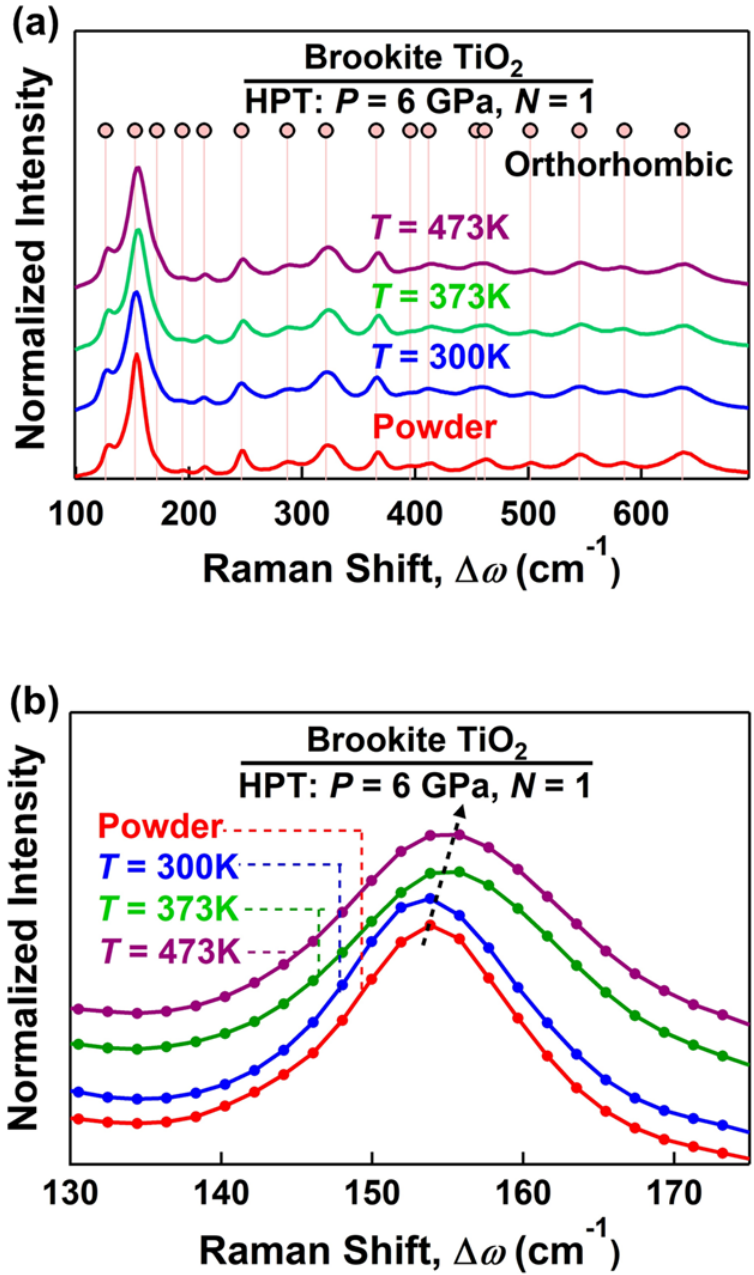

Fig. 4. (a) Raman spectra and (b) magnified Raman profile for the peak around 155 cm$^{-1}$ for brookite samples before and after HPT at 300, 373 and 473 K.

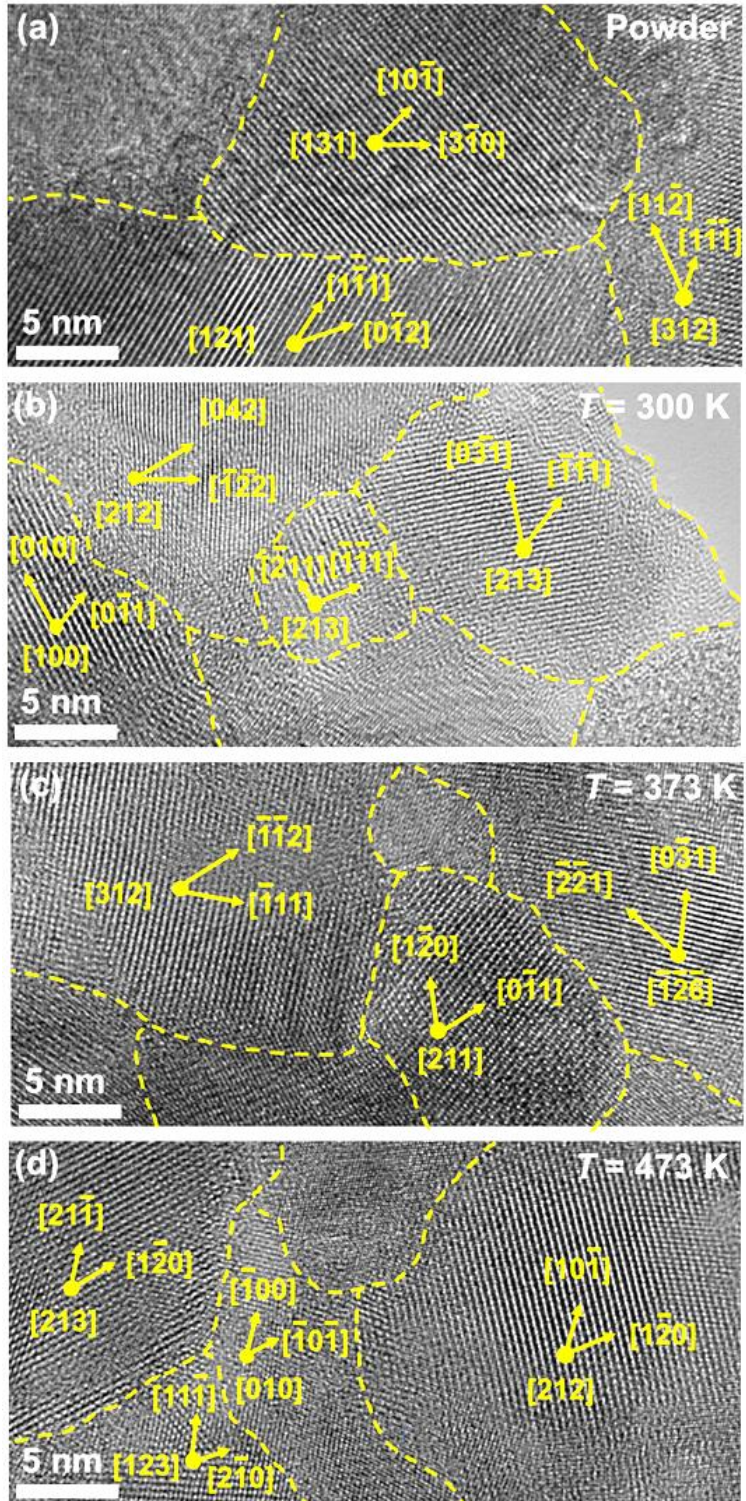

Fig. 5. TEM high-resolution images of nanograins for brookite samples (a) before HPT and (b) after HPT at 300 K, (c) 373 K and (d) 473 K.

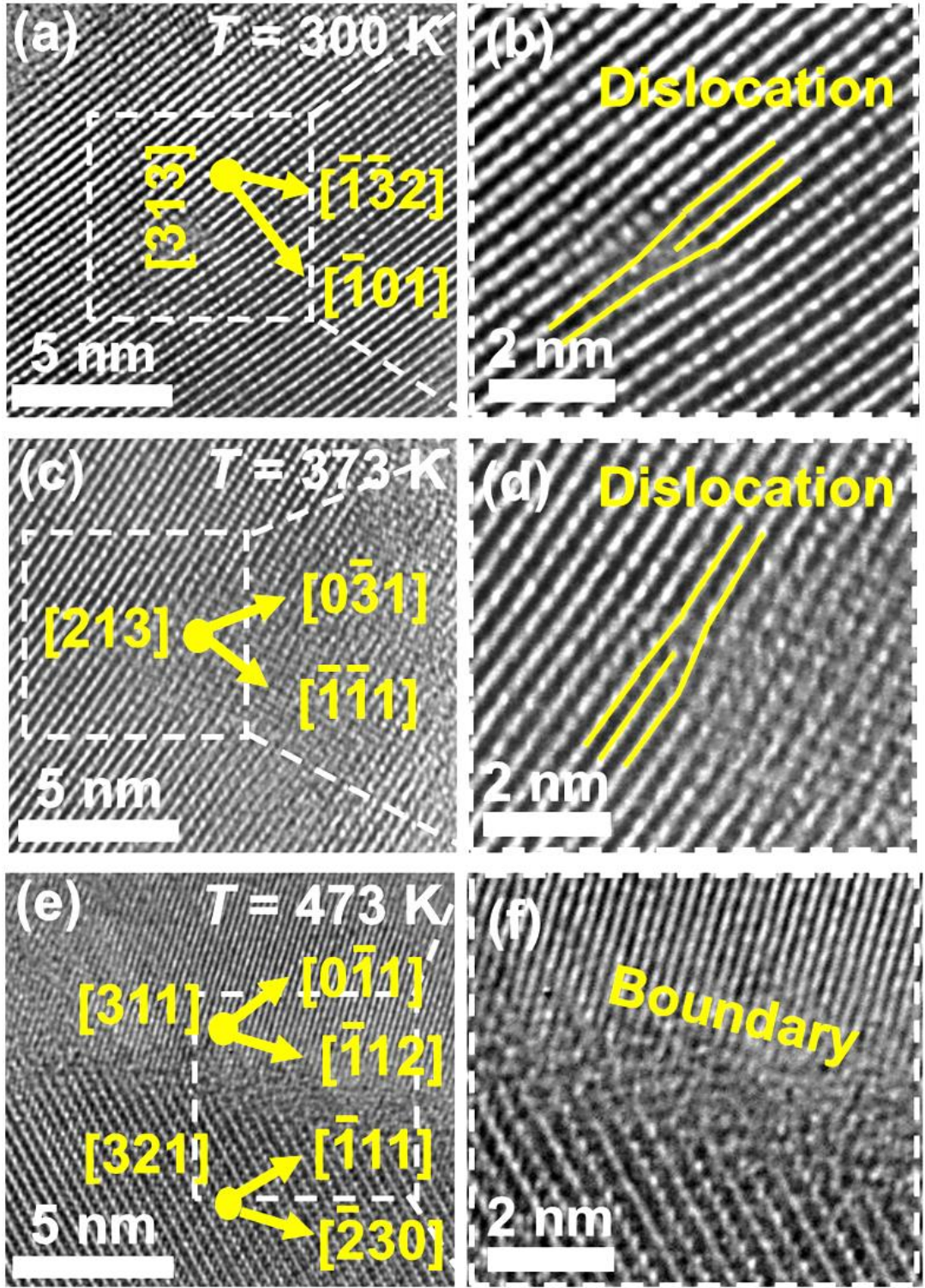

Fig. 6. TEM high-resolution images of dislocations and grain boundaries for brookite samples after HPT at (a,b) 300, (c,d) 373 and (e,f) 473 K, respectively.

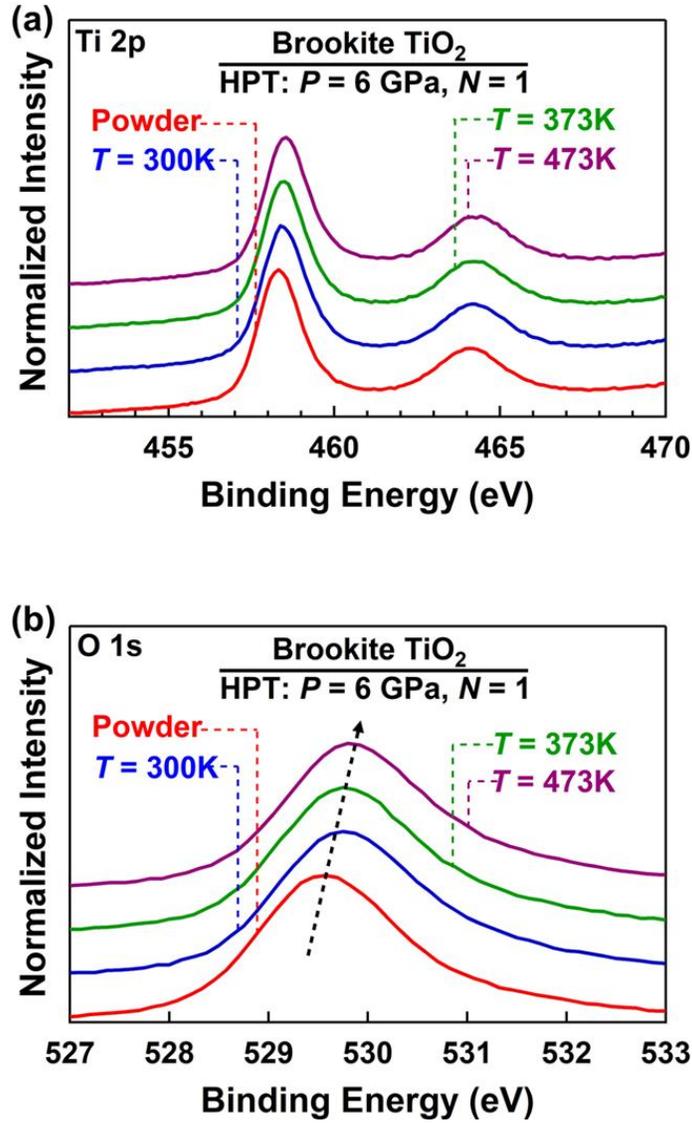

Fig. 7. XPS spectra of (a) Ti 2p and (b) O 1s for brookite samples before and after HPT at 300, 373 and 473 K.

Fig. 8a shows the photoluminescence spectra to evaluate the recombination rate of electrons and holes. The appearance of a peak around 400 nm (3.1 eV) for the initial powder corresponds to band-to-band electron-hole recombination [7] which disappears after HPT processing indicating the significant suppression of this kind of recombination after HPT. Another peak that appears at around 520 nm, corresponding to the recombination of electrons and holes on defects [7], becomes also less intense after HPT processing. Fig. 8b demonstrates the EPR spectra of four samples. EPR spectroscopy for brookite $TiO_2$ usually shows the existence of $Ti^{3+}$ on the surface and in bulk. The figure shows the appearance of two pair peaks at $g$ factors around 1.994 and 2.018. The first peak that appeared at $g = 1.994$ corresponds to the formation of $Ti^{3+}$ in the bulk and the second one with $g = 2.018$ corresponds to the existence of inherent $Ti^{3+}$ on the surface of brookite [5,37]. The peak appearing at $g = 1.994$ is weak for the powder and then empowers in HPT samples due to the formation of large fractions of oxygen vacancies [5]. The intensity of the EPR peak for bulk defects is more significant than that of surface defects due to two reasons: (i) HPT as a severe plastic deformation technique introduces defects mainly in bulk [38], although the samples after HPT are crushed to have defects every including bulk and surface; and (ii) defects in the bulk are naturally higher than on the surface due to a larger volume of bulk compared to the surface. The black brookite produced by HPT processing at 473 K exhibits the most intense peaks for both surface and bulk oxygen vacancies.

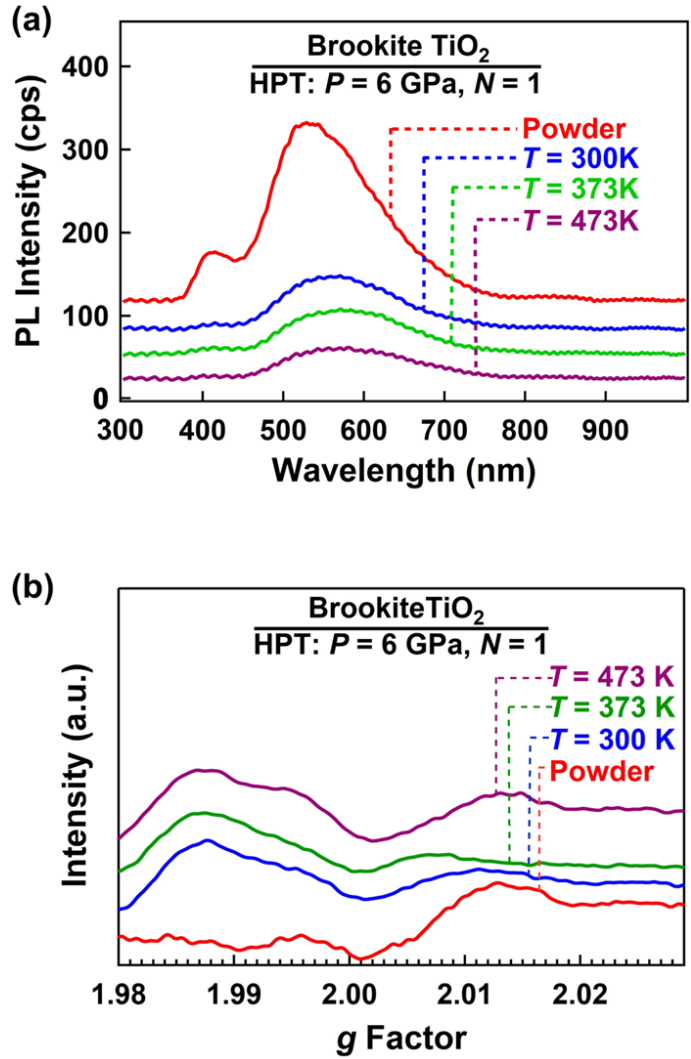

Fig. 8. (a) Photoluminescence and (b) EPR spectra for brookite samples before and after HPT at 300, 373 and 473 K.

Fig. 9a shows the UV-vis spectra of all samples before and after the HPT process. The light absorbance of brookite $TiO_2$ increases by HPT processing, and the sample processed at 473 K has the highest light absorbance in the visible light region. The light absorbance data agree with the change of sample colors from white to light grey, dark grey and black, respectively. The Kubelka-Munk theory was used to calculate the bandgap values as shown in Fig. 9b, where $\alpha$, h and $v$ are light absorption, Planck's constant and photon frequency, respectively. Bandgap values decrease from 3.1 eV for the powder to 2.8 eV for the samples processed by HPT. Fig. 9c shows the XPS spectra to calculate the top of the valence band. Measured values for the initial brookite and HPT samples processed at 300, 373 and 473 K are 2.0, 2.4, 2.4 and 2.2 eV versus NHE level, respectively. By considering the bandgap and the top of the valence band values, the bottom of the conduction band can be calculated by adding the bandgap value to the top of the valence band. Values for the bottom of the conduction band are -1.1, -0.4, -0.4 and -0.6 eV versus NHE for powder and HPT samples processed at 300, 373 and 473 K, respectively. If these values are converted to the values versus vacuum level, the electronic band structures shown in Fig. 9d can be achieved. A comparison between these band structures with chemical potentials for various $CO_2$ conversion reactions confirms that all four samples can support all $CO_2$ conversion reactions [8-17].

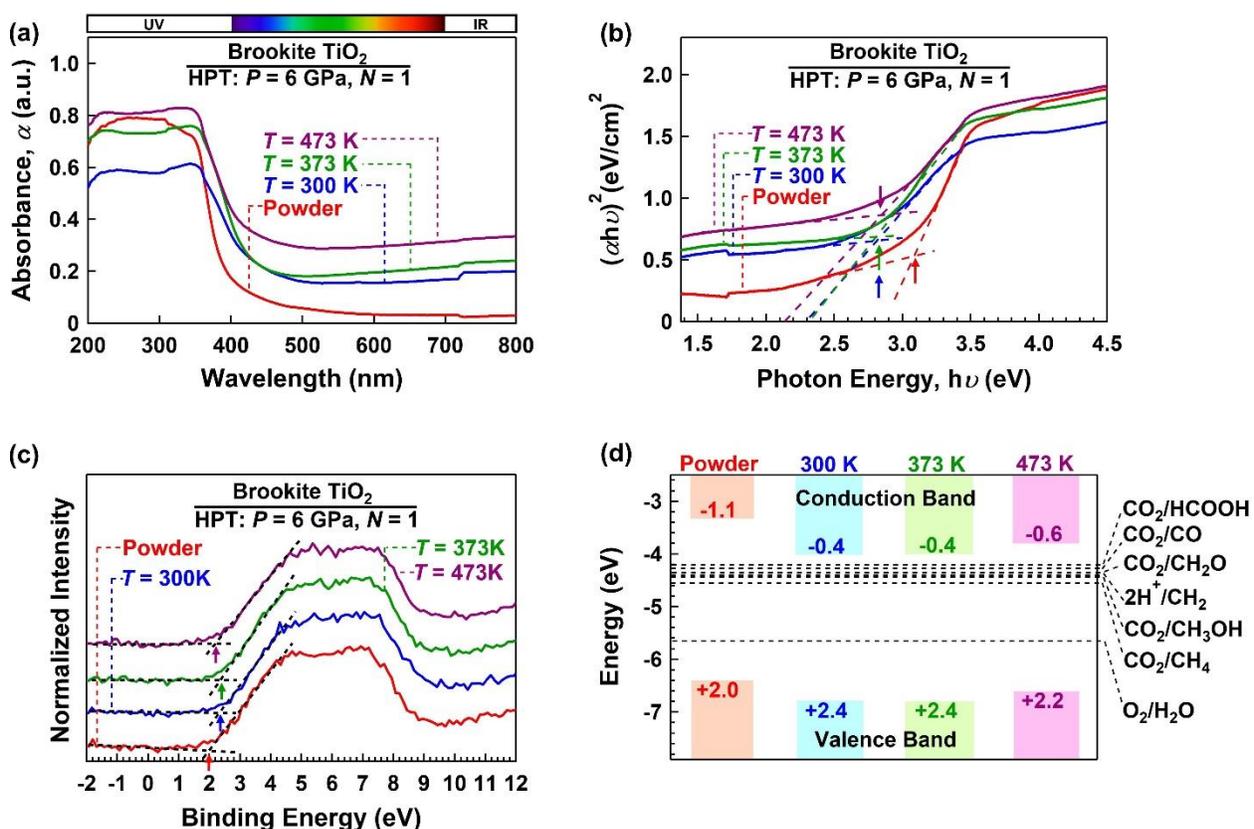

Fig. 9. (a) UV-vis spectra, (b) Kubelka-Munk plot for bandgap calculation, (c) XPS spectra for top of valence band calculation and (d) electronic band structure for brookite samples before and after HPT at 300, 373 and 473 K. Numbers written in (d) are values of top of valence band and bottom of conduction band versus NHE.

Fig. 10a shows the photocatalytic $CO_2$ conversion results for brookite samples before and after HPT processing. The samples processed by HPT at 300 and 373 K have activities reasonably similar to the initial power for both $H_2$ and CO production, while the black sample processed at 473 K shows over three times higher efficiency than the initial powder. The average concentration of produced CO for the initial powder is 2.76 µmolh$^{-1}$m$^{-2}$. For the black sample processed at 473

K, the average value increases and reaches $13.00 \pm 2.56$ µmolh$^{-1}$m$^{-2}$. The average H$_2$ production rate for the powder is $1.88 \pm 0.04$ µmolh$^{-1}$m$^{-2}$ and increases to $10.56 \pm 1.27$ µmolh$^{-1}$m$^{-2}$ for the black brookite processed at 473 K. Although the optimum processing temperature needs to be examined by modification of the HPT facility to increase the temperature to over 473 K, the present data suggests that a higher processing temperature is desirable to achieve a higher photocatalytic efficiency. The higher efficiency of the HPT sample processed at 473 K for CO$_2$ conversion can be attributed to the formation of oxygen vacancies as active sites for CO$_2$ adsorption and conversion, the lower recombination rate of electrons and holes, higher light absorbance and a narrower bandgap. To confirm the stability of photocatalysts after reaction, XRD analysis was performed after photocatalysis, as shown in Fig. 10c. All samples have the same XRD spectra before and after the photocatalytic test, indicating the stability of these brookite photocatalysts. Here it should be noted that the CO production rate of $13.00 \pm 2.56$ µmolh$^{-1}$m$^{-2}$ is much larger than the photocatalytic CO$_2$ conversion rate on the P25 TiO$_2$ benchmark photocatalyst when the same experimental setting is exactly used [23]. Moreover, the activity of black brookite is higher than HPT-processed anatase which exhibits an HPT-induced transformation from anatase to a mixture of defective high-pressure columbite phase and anatase phase [24].

## 5. Discussion

Black brookite with high photocatalytic activity for CO$_2$ conversion was produced by HPT processing in this study. Defective black brookite was developed in this study because earlier works confirmed the high activity of defective TiO$_2$ polymorphs for CO$_2$ conversion [18,19,39], and HPT was selected for processing because earlier works confirmed the potential of the method in the synthesis of various defective oxides and oxynitrides [40-42]. Although the material was produced by the HPT method, such a black oxide can be also produced by chemical methods that were used earlier to produce different black catalysts [18,19]. The formation of black brookite was successful in solving two drawbacks of brookite including a high recombination rate of electrons and holes and low light absorbance due to a large bandgap [5-8]. All observed features can be considered as a result of HPT-induced defect introduction such as oxygen vacancies, Ti$^{3+}$ radicals, dislocations and nanograin boundaries as effective sites to accelerate the electron-hole separation and migration [28,43-45].

Formation of oxygen vacancies in black brookite, experimentally proved by XRD, Raman spectroscopy, XPS and EPR, has the most significant impact on photocatalytic CO$_2$ conversion because vacancies can act as active sites to absorb and activate CO$_2$ on the surface of photocatalysts [46-49]. Vacancies can also act as shallow traps for electrons to enhance electron-hole separation and migration to the surface [46-49]. Moreover, oxygen vacancies can generate shallow traps to enhance the light absorbance and decrease the optical bandgap [46-49]. To confirm the effect of oxygen vacancies on light absorbance and CO$_2$ adsorption, DFT calculations were employed in this study.

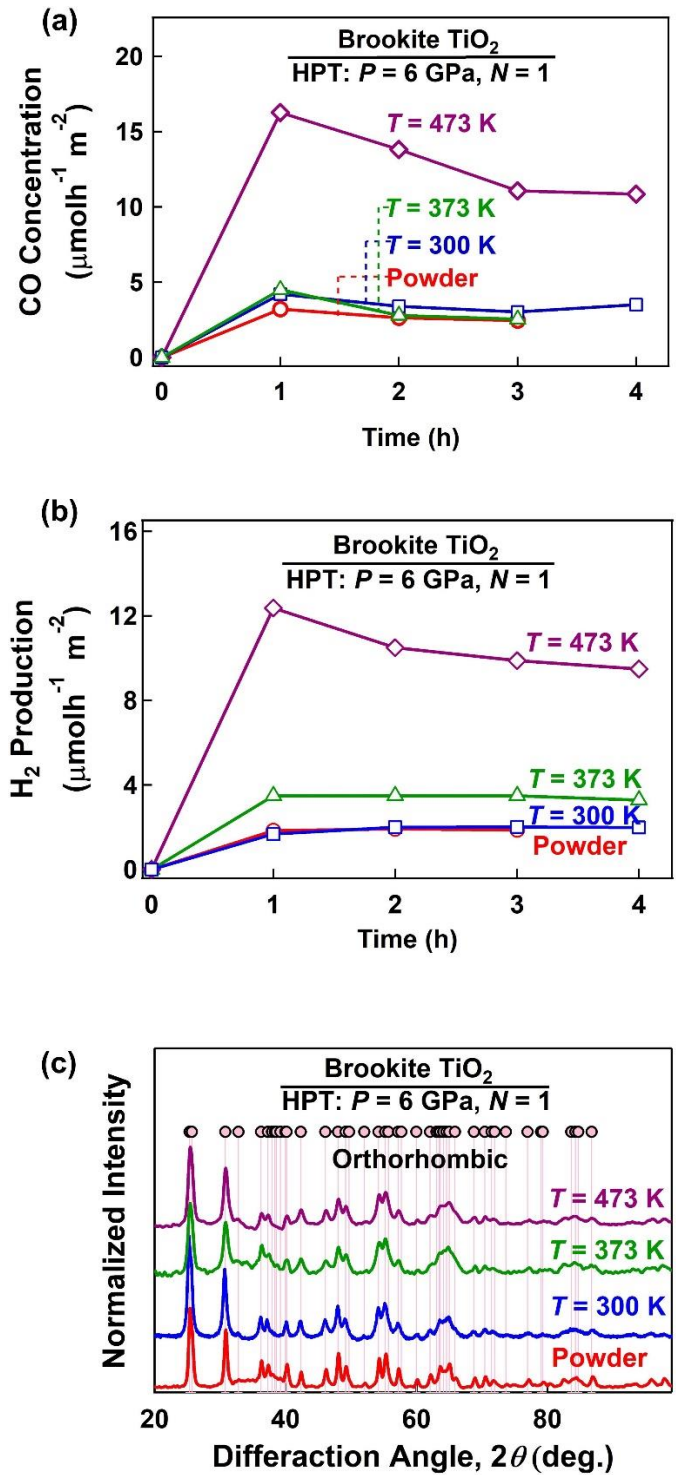

Fig. 10. Photocatalytic (a) $CO_2$ conversion and (b) $H_2$ production versus reaction time and (c) XRD profile after photocatalysis for brookite samples before and after HPT at 300, 373 and 473 K.

Fig. 11 compares the density of states for pristine and oxygen-deficient brookite supercells. For both cases, oxygen 2p orbitals are the main contributor to the valence band, while titanium 3d orbitals are the main contributor to the conduction band. The calculated bandgap for pristine brookite is 3.22 eV, but it decreases to 2.70 eV for the sample with 6% oxygen vacancies due to merging the defect states with the valence and conduction bands. Such a reduction in the optical bandgap confirms that the presence of oxygen vacancies in brookite can enhance its light absorbance and accordingly improve its photocatalytic activity. The bandgap of 3.22 eV calculated for the pristine structure coincides with the measured value in Fig. 9b and with data reported in the literature [50-52].

Fig. 12 shows the adsorption of $CO_2$ to the surface of brookite (a) without and (b) with oxygen vacancies. The adsorption energy is -0.93 eV for pristine brookite and -10.81 eV for oxygen-deficient brookite. The negative adsorption energy confirms that brookite is basically a good candidate for photocatalytic $CO_2$ adsorption and conversion. The reduction of the adsorption energy with the generation of oxygen vacancies confirms that oxygen vacancies can enhance the adsorption of $CO_2$ to the surface of brookite. This observation is consistent with what was reported for $TiO_2$ photocatalysts where the presence of oxygen vacancies improves the electron transfer, thereby enhancing the surface interaction with $CO_2$ [3,50,53]. These results confirm the positive role of oxygen vacancies in black brookite as active sites for $CO_2$ adsorption and subsequently photocatalytic reaction.

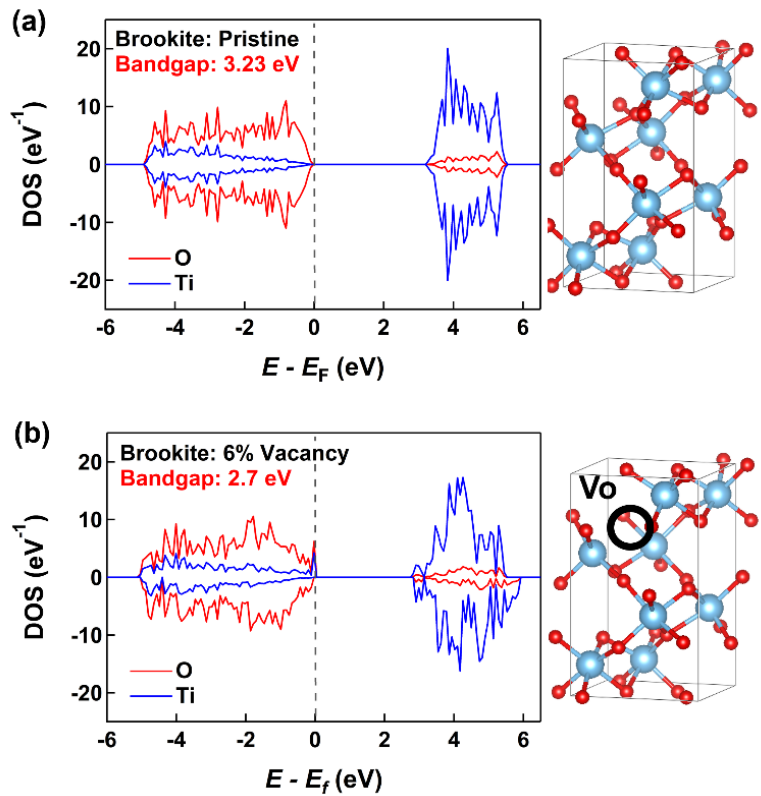

Fig. 11. Density of states (DOS) and corresponding supercells used in DFT calculations for (a) pristine brookite and (b) brookite with 6% oxygen vacancy.

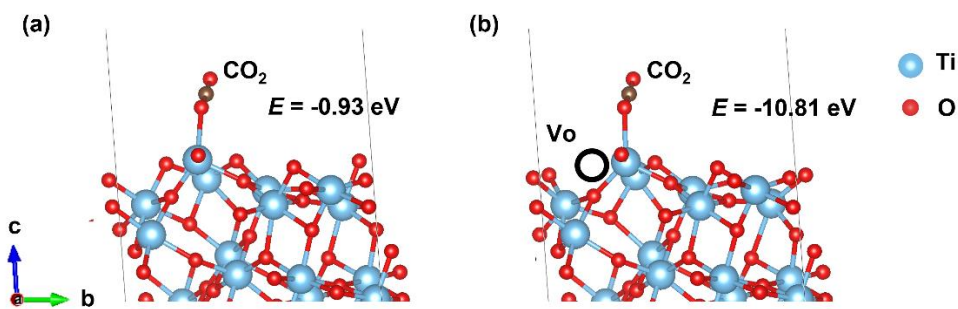

Fig. 12. Adsorption of $CO_2$ on (111) atomic plane and corresponding adsorption energy calculated by DFT calculations for (a) pristine brookite and (b) brookite with 6% oxygen vacancy.

Finally, it should be noted that the improvement of photocatalytic activity by oxygen vacancies and the formation of black brookite has the potential to be combined with other strategies employed for the improvement of activity of brookite such as the formation of heterojunctions [8-13], producing nanosheets [5], introducing nanorods [14], surface decoration by co-catalysts [15], shape controlling by amino acids [16] and doping [17]. It should be noted that although this work reports the application of black brookite for photocatalytic $CO_2$ conversion, the methodology used in this study can be basically used for producing any other defective black catalysts.

## 6. Conclusions

The first application of black brookite for photocatalytic $CO_2$ conversion was reported in this study. The high-pressure torsion method produced the black brookite with simultaneous generation of defects such as oxygen vacancies, $Ti^{3+}$ radicals, dislocations and nanograin boundaries. The synthesized black brookite had higher light absorbance, narrower bandgap and lower recombination rate of electrons and holes compared to conventional white brookite powders. The material converts $CO_2$ to CO and produces $H_2$ under UV irradiation with an activity better than initial white brookite. First-principles calculations suggest that such a high activity is mainly due to the effect of oxygen vacancies on enhancing the light absorbance and $CO_2$ absorbance, a fact that can be employed to produce other active photocatalysts.

**Acknowledgments**

This study is supported partly by Mitsui Chemicals, Inc., Japan, partly by Hosokawa Powder Technology Foundation, Japan, and partly through Grants-in-Aid from the Japan Society for the Promotion of Science (JSPS), Japan (JP19H05176 & JP21H00150). The author JHJ acknowledges a scholarship from the Q-Energy Innovator Fellowship of Kyushu University.

**Supplementary Material**

Supplementary material includes Fig. S1, Fig. S2 and Fig. S3.

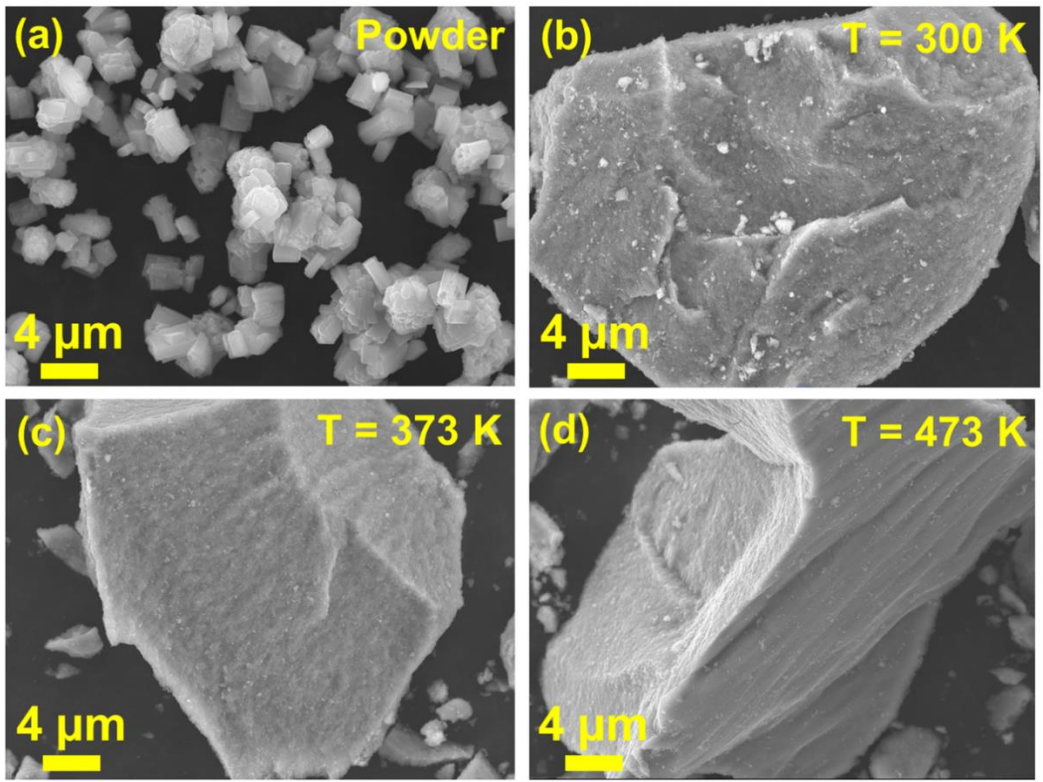

Fig. S1. SEM micrograph of brookite samples (a) before HPT and at (b) after HPT at 300 K, (c) 373 K and (d) 473 K.

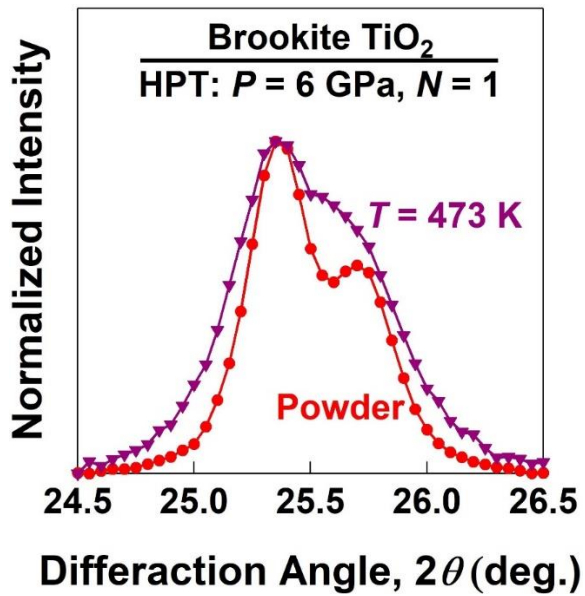

Fig. S2. Magnified XRD spectra for brookite samples after HPT at 473 K in comparison with initial powder to show peak broadening.

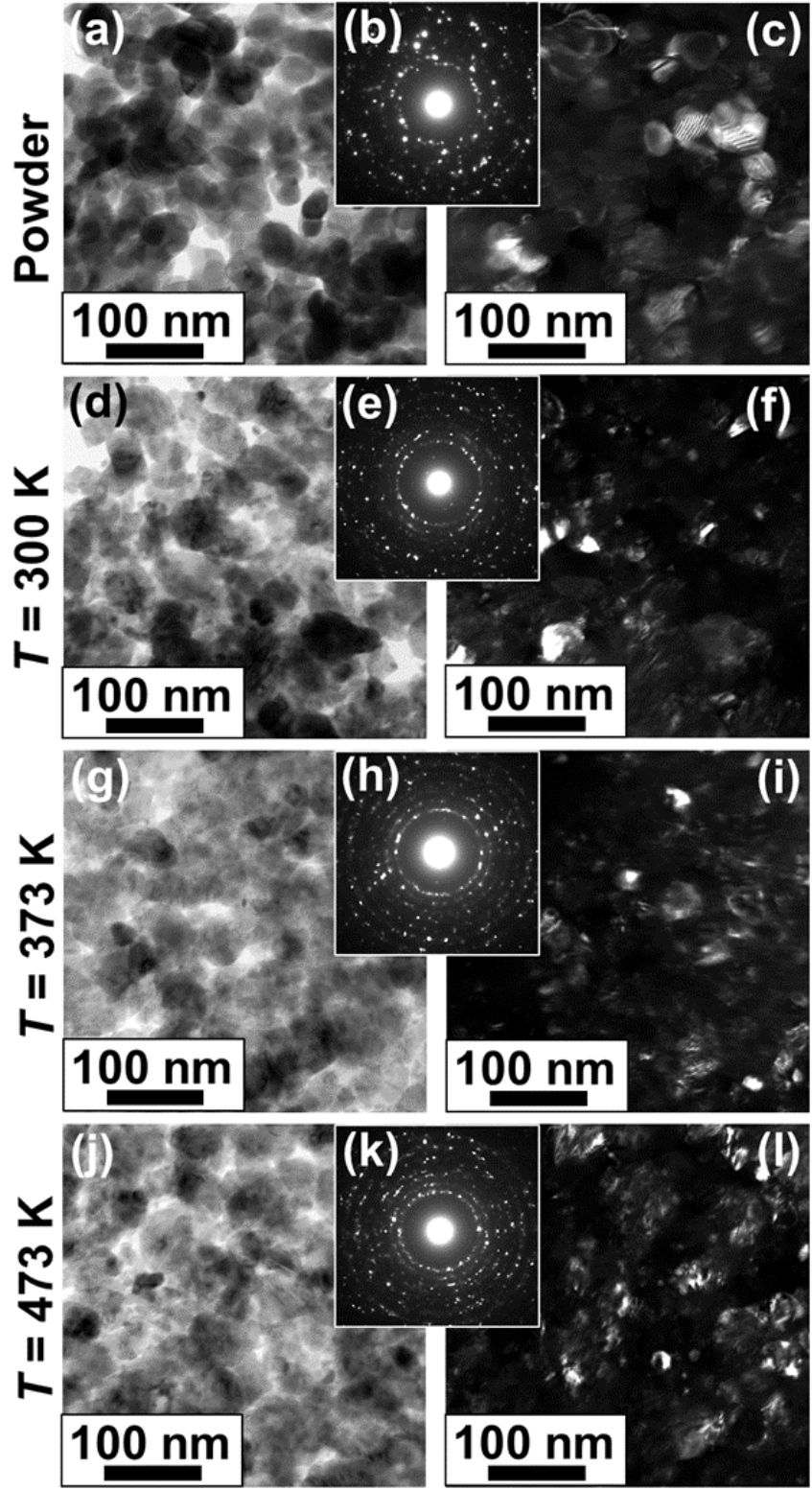

Fig. S3. Bright field, SAED and dark field TEM images for brookite samples (a-c) before HPT and (d-f) after HPT at 300 K, (g-i) 373 K and (j-l) 473 K.